\documentclass{article}
\usepackage{ijcai19}

\usepackage[round]{natbib} 

\usepackage{times}
\usepackage{soul}
\usepackage{url}
\usepackage[hidelinks]{hyperref}
\usepackage[utf8]{inputenc}
\usepackage[small]{caption}
\usepackage{graphicx}
\usepackage{amsmath}
\usepackage{booktabs}
\usepackage{float}
\title{Conversational Help for Task Completion \\
and Feature Discovery in Personal Assistants\thanks{This work was presented at 2nd Workshop on Humanizing AI (HAI) at IJCAI'19 in Macao, China.}}

\author{
Madan Gopal Jhawar\and
Vipindeep Vangala\and
Nishchay Sharma\and \\
Ankur Hayatnagarkar\and
Mansi Saxena\and
Swati Valecha\\
\affiliations
Microsoft India Development Center
\emails
\{majhawar, vipinv, kusharma, anhay, masaxe, swativ\}@microsoft.com
}

\begin{document}

\maketitle

\begin{abstract}
Intelligent Personal Assistants (IPAs) have become widely popular in recent times. Most of the commercial IPAs today support a wide range of skills including \emph{Alarms, Reminders, Weather Updates, Music, News, Factual Questioning-Answering,} etc. The list grows every day, making it difficult to remember the command structures needed to execute various tasks. An IPA must have the ability to communicate information about supported skills and direct users towards the right commands needed to execute them.

Users interact with personal assistants in natural language. A query is defined to be a \emph{Help Query} if it seeks information about a personal assistant's capabilities, or asks for instructions to execute a task. In this paper, we propose an interactive system which identifies help queries and retrieves appropriate responses. Our system comprises of a C-BiLSTM based classifier, which is a fusion of Convolutional  Neural  Networks (CNN)  and  Bi-directional  LSTM (BiLSTM) architectures, to detect help queries and a semantic Approximate Nearest Neighbours (ANN) module to map the query to an appropriate predefined response. Evaluation of our system on real-world queries from a commercial IPA and a detailed comparison with popular traditional machine learning and deep learning based models reveal that our system outperforms other approaches and returns relevant responses for help queries.
\end{abstract}

\section{Introduction}
In the era of Artificial Intelligence (AI), IPAs have gained a lot of attention and found their way into millions of homes. \emph{Google Now} from Google, \emph{Siri} from Apple, \emph{Alexa} from Amazon and \emph{Cortana} from Microsoft are the four major commercial IPAs in market today. All of them support a wide range of tasks including setting up alarms, getting weather updates, tracking flights, playing music, etc, and are adding many other skills and scenarios every day. 

However, skill expansion brings with it multiple unique challenges. First of all, \emph{feature discovery} becomes a critical factor. As the system evolves continuously, it is difficult for the users to stay updated with the latest and upcoming skills. The most common and widely used way of solving this problem is to have a help page which states all the existing and new skills. Other approaches include building a FAQ bot which can respond to factual user queries. However, these are very naive and non-intuitive approaches. With most of the IPAs having a voice-enabled interface, a more interactive ecosystem is needed where users interact with the personal assistant in natural language to discover its new features. Secondly, every skill has its own functionality and a set commands that it supports. For example, a \emph{``Weather''} skill which provides users with the weather updates and forecast from any part of the world is reasonably different from an \emph{``Alarm''} skill which allows users to \emph{create}, \emph{delete} or \emph{snooze} alarms. These skills serve very different purposes and have their own independent set of commands. Therefore, IPAs need to have a system to guide users towards the right command structures needed to execute any task and educate them about the overall utility of a skill. Lastly, for a smooth, fluent and satisfactory user experience, a personal assistant should assist the users in completing the intended tasks and guide them to use the right set of commands in a seamless manner.

A query is defined to be a \emph{Help} if it's intent is one of the following -- a) asks for the IPA's overall capabilities b) seeks information about a particular skill c) directly seeks help for executing a specific task d) questions about instructions required to complete a task e) showing interest in any task. Based on the intent, help queries can broadly be divided into two categories -- \emph{skill help} and \emph{generic help}. Queries which seek information about the overall capabilities of the system, \emph{intent a)}, fall under the generic help category. On the other hand, queries that seek information about a particular skill of the system, \emph{intents b)-e)}, are defined as skill/task help queries. Table \ref{tab:sampleQueries} tabulates some sample queries with each of the above defined intents.

\begin{table}
\caption{Possible \emph{Help} intents and sample queries}
\label{tab:sampleQueries}  
\centering
\begin{tabular}{ lll } 
\hline
\textbf{Intent}  & \textbf{Sample} & \textbf{Query}\\
& \textbf{Query} & \textbf{Type} \\
\hline
Overall capabilities  & What can & Generic  \\
of the IPA &  you do? &  \\
\hline
Information about  & Do you have & Skill   \\
a particular skill & flight tracking? &   \\
\hline
Direct help  & Help me to & Skill  \\ 
for a task & setup an alarm. &  \\
\hline
 Questions on task  & Tell me how to & Skill   \\ 
 instructions & connect via bluetooth? &   \\ 
\hline
Interest in a task & I would like & Skill \\
& to play music. &  \\
\hline
\end{tabular}
\end{table}

In this paper, we propose a system which identifies \emph{help} queries and provides appropriate responses. The system is made up of two modules. The first module uses \emph{C-BiLSTM} deep learning architecture, which is a fusion of Convolutional Neural Networks (CNN) and Bi-directional LSTMs (BiLSTM), to detect help queries. Given an input query, C-BiLSTM first passes it through a convolution layer to learn various important local features, which in turn are passed through a BiLSTM layer which encodes various sequential information in the query by making a forward as well as backward pass over the query features. The deep features, thus obtained, are fed into a Fully Connected (FC) layer to predict whether the query belongs to help class or not. We have also compared this model with various hand-crafted features based traditional machine learning models, and deep learning based individual CNN, LSTM and BiLSTM models. Results on a real-world help query dataset reveals that C-BiLSTM outperforms other models.  

To map the identified  help query to one of the predefined responses,  we leverage the labelled training data to find the queries which are semantically  similar  to  the  input  query. To achieve this in real-time, we use KDTrees-based Approximated Nearest Neighbours (ANN) (\cite{panigrahy2008improved}) approach to find \emph{top-k} similar help queries using their DSSM (\cite{huang2013learning}) feature vectors. We return the response of the closest matched query if the corresponding response is in majority and the similarity score is above a certain threshold. 

In summary, the following are our main contributions in this paper:
\begin{itemize}
    \item We introduce the research problem of conversational help systems for task completion and feature discovery in personal assistants.
    \item We propose an interactive conversational help system which detects \emph{help} queries and provides appropriate responses.
    \item We evaluated and compared various latest deep learning techniques including CNN, LSTM, BiLSTM and C-BiLSTM for classifying help queries.
    \item We used KDTree-based ANNs on DSSM semantic query features to find similar queries in real-time and use them to map the input query to an appropriate response.
\end{itemize}

The rest of the paper is organized as follows. Section \ref{relatedwork} presents previous works done in the field of personal assistants and conversational task completion systems. Section \ref{helpSystem} describes our approach for building the system. Section \ref{expNresults} describes the experimental setup, details of the dataset used and various quantitative and qualitative results. Section \ref{conclusions} concludes the paper.

\section{Related Work} \label{relatedwork}
Ever since the evolution of personal assistants, various studies pertaining to user understanding, context aware conversations and recommendations, etc., have been carried out in literature. \cite{guha2015user} proposed a system that uses web search history to identify coherent contexts that correspond to tasks, interests, and habits of the signed-in \emph{Google Now} users. They look at several months of history in order to identify not just short-term tasks, but also long-term interests and habits of the users. \cite{elwany2014enhancing} analyzed millions of user queries from \textit{Cortana} to build a machine learning system capable of classifying user queries into two classes -- a class of queries that are addressable by Cortana with high user satisfaction, and a class of queries that are not. \cite{kumar2017incomplete} proposed a retrieval-based sequence to sequence learning system that generates complete (intended) question for an incomplete follow-up question, given the conversation context, in \textit{Siri}. \cite{lopatovska2018talk} examined user interactions with \textit{Alexa}, and focused on the types of tasks requested of Alexa, the variables that affect user behaviors with Alexa, and Alexa’s alternatives.

In some other works, interactive systems have been deployed to provide conversational assistance in completing tasks. \cite{thompson2004personalized} proposed a system that treats item selection as an interactive, conversational process, with the program inquiring about item attributes and the user responding. \cite{levinson1980conversational} designed a conversational-mode, speech-understanding system which enables its user to make airline reservations and obtain timetable information through a spoken dialog. \cite{janarthanam2013multithreaded} demonstrated a conversational interface that assists pedestrian users in navigating within urban environments and acquiring
tourist information by combining spoken dialogue system, question-answering, and geographic information system technologies. 
Some progress have been done in building question answering system from FAQ (Frequently Asked Questions) files. \cite{burke1997question} used a purely lexical metric of
similarity between query and document and a semantic knowledge base (WordNet) to match question and answer from FAQ files. Calculation of question similarity is a key problem in FAQ answering which was solved using question vector similarity based on WordNet (\cite{song2007question}) and using similarity between answers in the archive to estimate probabilities for a translation-based retrieval model (\cite{Jeon:2005:FSQ:1099554.1099572}).

However, to the best of our knowledge, there is no prior work related to conversational help systems for task completion and feature discovery in personal assistants. In this paper, we propose a two-level system where we first classify help queries, and then use semantic similarity between queries to map the input query to an appropriate response.

\begin{figure*}
\caption{Workflow of the Help system. DSSM word vectors of the input query is sequentially passed through CNN, BiLSTM and Fully Connected layers to classify the query into help or not help categories. DSSM semantic similarity based ANNs of the query are used to map it to an appropriate response.}
\label{fig:workflow}
\centering
\includegraphics[scale=0.7]{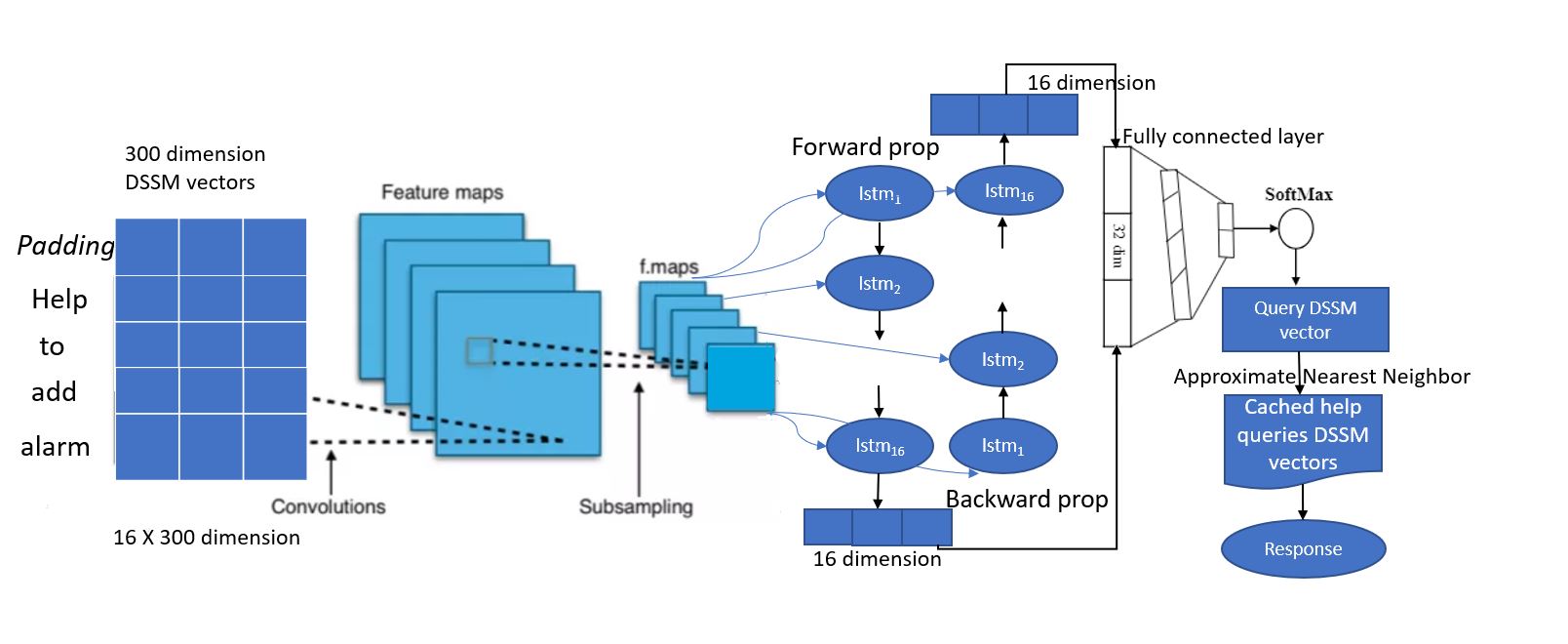}
\end{figure*}

Over the last decade, deep learning based models have achieved great success in many of the NLP tasks, including sentence classification. \cite{kim2014convolutional} showed that a simple CNN with little hyper-parameter tuning and static vectors achieves excellent results, and improved upon the state-of-the-art on 4 bench-marking tasks including sentiment analysis and question classification. On the other hand, RNNs with Long Short-Term Memory units \cite{mikolov2012statistical,chung2014empirical,tai2015improved} are
effective networks to process sequential data, which analyze a text word by word and stores the semantics of all the previous text in a fixed-sized hidden state. In this way, LSTM can better capture the contextual information and semantics of long texts. Moreover, bidirectional RNNs (\cite{schuster1997bidirectional}) processes the sequence both forward and backward, naturally, a better semantic representation can usually be obtained than unidirectional RNNs. \cite{zhou2015c} combined CNN and LSTM architectures for sentence representation and text classification. Recently, \cite{yenala2017convolutional} combined the strengths of CNNs and BiLSTMs and proposed \emph{C-BiLSTM} to classify inappropriate queries on real-world web search dataset. They evaluated other traditional as well as deep learning based classifiers and showed that C-BiLSTM performed better than individual CNN, LSTM and BiSLTM models. 

\cite{huang2013learning} proposed DSSM, a deep neural network modeling technique for representing text strings (sentences, queries, predicates, entity mentions, etc.) in a continuous semantic space. DSSM can be used to develop latent semantic models that project multiple entities into a common low-dimensional semantic space for a variety of machine learning tasks such as ranking and classification. For example, for finding similar queries, the relevance between two queries can be readily computed as the cosine similarity between them in that space. The DSSM similarity score is found to be better than many traditional features including TF-IDF, BM25, Word Translation Model (WTM), LSA, PLSA, and LDA, etc. (\cite {yebuilding,bogdanova2017sesa}), and has been widely used for various semantic similarity based tasks (\cite{ye2016enhancing,palangi2014semantic}). For the task of fetching appropriate response, we used DSSM based semantic similarity between queries.

\section{Conversational Help System} \label{helpSystem}

The architecture of the conversational help system is depicted in Figure \ref{fig:workflow}. The system is divided into two modules -- \emph{Help Query Classification} and \emph{Response Mapping}. 

A fusion model which combines the strengths of convolutional neural networks and bidirectional LSTMs is used to detect help queries. The model takes the word embedding matrix of the query as input and passes it through three sequential layers -- a) Convolutional (CONV) Layer, b) Bi-directional LSTM (BLSTM) Layer, and c) Fully Connected (FC) Layer. The CONV layer first project the input query embedding matrix into a lower-dimensional feature representation. The BLSTM layer encodes sequential information by making forward and backward passes over the query features. The FC layer finally models the query features and outputs whether it is a help query or not. 

Once the input query is identified as a help query, a distributed query similarity model is used to fetch the appropriate response. We use deep semantic features of the input query to find \emph{top-k} similar help queries from the training data, which have been cached offline. We use majority voting to pick the most appropriate response suitable for the input query.


\subsection{Help Query Classification}
In this section, we will explain the working of help classifier in detail.

\subsubsection{Query Normalization} 
Real-world user queries come with a lot of noise in the form of spelling mistakes, incorrect sentences, use of slang words and phrases, short-forms, etc. Therefore, we have carried out several domain-specific as well as standard normalization steps. We convert the input query to lower case and remove stop words and punctuation marks, followed by lemmatization. We translate more than 50 most frequently used slang words and phrases, such as wanna, gimme, etc., to their English counterparts. Since \emph{Alarms}, \emph{Reminders} and \emph{Timers} are among the supported skills, we observed that a good portion of the queries contained time stamps. Some of the examples of such queries are -- \textit{``how to set an alarm for 5:00pm'', ``help me to cancel the 10:00pm reminder for tomorrow''}, etc. Therefore, we used regular expressions to convert all the time stamps with the token \emph{``time\textunderscore stamp''}. Similarly, \emph{Music} and \emph{Radio} are other top skills. Hence, we mined top-300 music genres and replaced them with the token \emph{``music\textunderscore genre''}.

\subsubsection{Query Embedding}
DSSM (\cite{huang2013learning}) has been trained on real-user queries which makes it our first choice for word embedding. Furthermore, it uses character-trigrams which enables it to generate embedding for misspelled and unseen words. Therefore, DSSM word embedding is used to project every word in the input query into a 300-dimensional feature space, forming the query matrix. As the length of the input queries could vary, we pad them with \emph{``unk''} tokens, representing unknown words, at the beginning to fix the input query length to \emph{maxlen} (15 in our case).

\subsubsection{The Convolutional Layer}
The representation of the input query is given by a matrix $Q \in R^{d*l}$, where $d$ is the dimensionality of word embedding and $l$ is the length of the input query (\emph{maxlen}). A 1-D convolution layer is applied on the word embedding matrix $Q$ by convolving a filter $H \in R^{d*m}$ of length $m$, followed by a max-pooling layer to extract the important local features. An element-wise non-linear transformation using Rectified Linear Unit (ReLU) is applied after convolution operation to avoid saturation and improve learning.

\subsubsection{The BiLSTM Layer}
The fixed length feature vector obtained from the convolution layer is fed into the BLSTM layer. Bi-directional LSTMs make forward and backward passes over the input and capture relevant sequential patterns. The outputs of the forward and backward LSTM cells inside the BLSTM are concatenated to form a combined feature output encoding rich semantic features of the query. These features are passed onto a FC layer which models the various interactions between the features. The final softmax node in the FC layer outputs the probability of the query belonging to the help class.

\subsection{Response Fetching}
Cortana today supports a wide range of tasks, such as \emph{playing music}, \emph{creating alarms}, \emph{tracking packages}, etc. For each of such tasks, for which help has to be provided, we used human experts to get an \emph{appropriate response string} which can guide the users towards completing the task. 

After classifying an input query as a help query, we need to map it to one of the predefined responses. Since a help query can be framed in multiple ways, understanding the semantic meaning of the query becomes an important part. For example, to ask help for \emph{deleting an alarm}, queries can be framed as \emph{``help me to delete my alarm''}, or \emph{``how do I shut the alarm up''}, or \emph{``instructions on deleting an alarm''}, etc.

Therefore, we leverage the labelled training data to find \emph{top-k} queries which are semantically similar to the input query. We use cosine similarity between the DSSM features (DSSM model used is pretained with query-document pairs of a commercial search engine) of queries as the measure of semantic similarity between them. For this, in offline, we map all the help queries in our training data, a total of 50K queries, to their DSSM feature vectors. We store these feature vectors, along with their appropriate responses, in a cache so that they can be readily used in real-time. For the input query, we use its DSSM features to find its top-k (k=1 in our case) ANNs, using KDTrees (\cite{panigrahy2008improved}), from cached feature vectors. The ANN algorithm finds approximately similar queries with a very low latency. If the similarity between the closest query and the input query is above a certain threshold, we return the response corresponding to the matched query.

\section{Experiments and Results} \label{expNresults}
In this section, we describe the dataset used, the quantitative results of various experiments we have carried out with feature and model selection for the help query classifier. Furthermore, we explain the challenges, experiments, and quantitative and qualitative results of the response mapping approaches.

\subsection{Dataset}
\begin{figure}
\caption{Distribution of queries from various skills in the dataset.}
\label{fig:skillcounts}
\centering
\includegraphics[width=85mm,height=55mm]{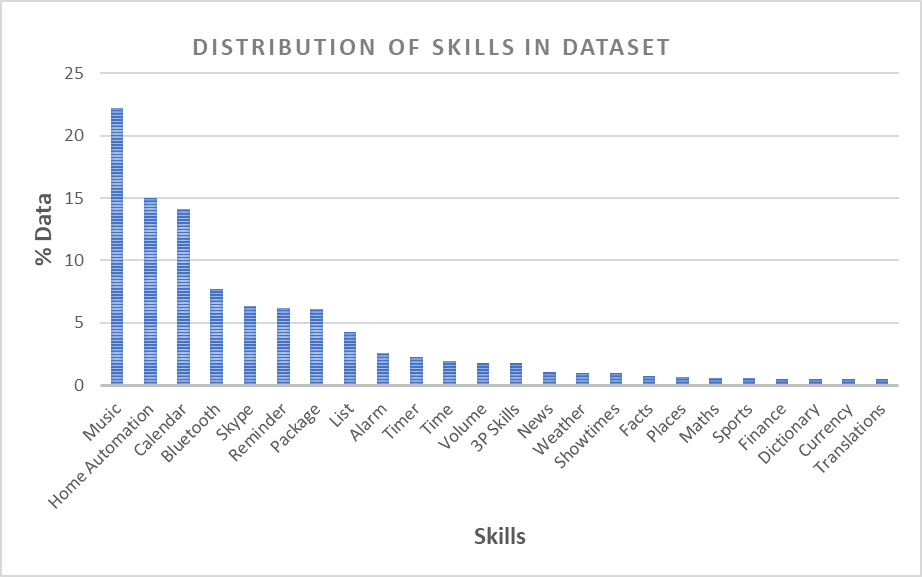}
\end{figure}

We have mined a dataset comprising of 200K unique queries from anonymized and privacy preserved Cortana logs of internal users. The data has been manually labeled by human judges into two categories -- help or not help. In case of help, they also tagged the queries with one of the predefined responses. To avoid human bias and judgment errors, each query was judged by three judges. The judges were given clear guidelines, along with sample query-label pairs, on how to classify the queries. A query was finally labeled help if majority of the judges labeled it as help. The dataset comprises of 24 different supported skills and varying tasks. The distribution of the skills in the dataset is depicted in Figure \ref{fig:skillcounts}. We have randomly divided the data into train, validation and test sets. The training set comprises of $80\%$ of queries, while the validation and test sets comprises of $5\%$ and $15\%$ of queries respectively.

\subsection{Help Query Classification}
For detecting help queries, we experimented with both hand-crafted features based as well as deep learning based classifiers. We used word-based unigram and bigram features to evaluate two widely used traditional machine learning algorithms -- Support Vector Machine (SVM) and Gradient Boosted Decision Trees (XGBoost). We found that XGBoost yielded slightly better recall at the cost of lower precision in classifying help queries. Among   the   deep   learning   based   classifiers, we evaluated the most popular classifiers including CNN based classifier, LSTM and BiLSTM  based  classifiers and compared them with C-BiLSTM model for detecting help queries. We used DSSM word embedding query matrix as the input features for all the deep learning based classifiers. The optimal hyper-parameters for various classifiers, as tuned on the Validation set, is tabulated in Table \ref{tab:parameters}.

\begin{table}
\caption{Optimal Hyper-Parameters for various deep learning models, as tuned on the Validation Set}
\label{tab:parameters} 
\centering
\begin{tabular}{l|l|l|l} 
Parameter & LSTM & BiLSTM & C-BiLSTM\\
\hline
Embd. Size  & 300 & 300 & 300\\
MaxLen  & 15 & 15 & 15\\
Batch Size  & 1024 & 1024 & 1024\\
Filter Size  & \textit{NA} & \textit{NA} & 128\\
LSTM cells  & 32 & 32 & 32\\
Optimizer  & Adagrad & Adagrad & Adagrad\\
Learning rate & 0.001 & 0.001 & 0.001\\
\end{tabular}
\end{table}

Table \ref{tab:quantResults} tabulates the precision, recall and F1-scores of the various models on the Test set. As it can be seen, all the deep learning based classifiers are able to outperform the traditional ngram-based classification algorithms. Also, while the LSTM and BiLSTM based classifiers produced comparable results, C-BiLSTM model yielded the best performance in detecting help queries. 

\begin{table}[h]
\caption{Classification performance of various models on the Test set. C-BiLSTM is found to outperform other models in detecting Help queries.}
\label{tab:quantResults}  
\centering
\begin{tabular}{ l|l|l|l } 
\textbf{Model}  & \textbf{Precision} & \textbf{Recall} & \textbf{F1 Score}\\
\hline
XGBoost & 0.844 & 0.809 & 0.824\\
SVM & 0.916 & 0.789 & 0.848\\
\hline
CNN & 0.914 & 0.807 & 0.857\\
LSTM & 0.906 & 0.810 & 0.855\\
BiLSTM & 0.92 & 0.819 & 0.867\\
C-BiLSTM & 0.952 & 0.826 & \textbf{0.885}\\
\end{tabular}
\end{table}

\subsection{Response Fetching}
Help queries, by definition, seek help about tasks. A task is defined as performing an \emph{action} of a \emph{skill}. For example, for the query \emph{``How to create an alarm?''}, \textit{alarm} is the skill and \textit{create} is the action. Similarly, for the query \emph{``Help me to play music.''}, \textit{music} is the skill and \textit{play} is the action. Therefore, we experimented with mapping help responses with supported \emph{action-skill} pairs. We expanded the actions list with their popular synonyms to increase the coverage. For example, queries \emph{``How to create an alarm?''} and \emph{``Can you set up an alarm?''} are mapped to the same action -- \emph{``create''}, as \emph{set up} is a synonym of \emph{create}.

To extract the action and skill, we used Parts-Of-Speech (POS) tags of the input query. We used the \textit{main verb} of the query as the action and the \textit{noun} as the skill. Thus, the query is mapped to the help response corresponding to the extracted action-skill pair. Upon evaluation on the test set, as there is no parameter tuning on the validation set, this approach yielded \emph{0.72 F1 score, with 0.89 precision and 0.61 recall scores}. A qualitative analysis of the approach and the results in tabulated in Table \ref{tab:posQualitative}.

\vspace{0.2cm}
\begin{table}[H]
\centering
\caption{Qualitative analysis of the POS-based response fetching module. This approach works well for simple help queries but fails for generic and unseen help queries.}
\label{tab:posQualitative}   
\begin{tabular}{ c|c|c }
 & \textbf{Action/} &  \\
\textbf{Query} & \textbf{Skill} & \textbf{Comments}\\
\hline
How to connect  & Connect/ & Mapped to correct \\
via bluetooth? & Bluetooth &  response. \\
\hline
Can you hook up & Hook up/ & \emph{Hook up} was not listed \\
via bluetooth? &  Bluetooth & as an action, resulting  \\
& & in recall gap. \\
\hline
Tell me the steps to & Sync/ & \emph{Sync} is not added \\
sync my smart tv. & TV  & to list of actions. \\
\hline
What can & Do/ & This approach doesn't \\
you do? & - &  work for \emph{generic} help  \\
& & queries, as there is \\
& & no skill mentioned. \\
\end{tabular}
\end{table}

As it can be seen, there are three major shortcomings of this approach.
First, it doesn't not support generic help queries as they do not have associated actions and skills. Second, this approach is not scalable as maintaining a map of verbs to actions requires constant human interventions. Also, it does not automatically scale up to new skills having different sets of actions. And third, while the approach works well for simple help queries, it lacks semantic query understanding required for complex queries.

Therefore, we experimented with DSSM-based semantic similarity between input query and the existing help queries. The input query is mapped to the response corresponding to the best match. This approach helps us in overcoming all the three challenges:
\begin{itemize}
    \item Training data comprises of both generic as well as task help queries, and the input query is compared with all of them.
    \item Instead of maintaining and updating an ever-lasting map of verbs to actions, we seed the existing and new skills with a few sample queries and the semantic similarity module maps the input query to an existing query. The new queries are added to the training data and corpus expands over time, further improving the performance.
    \item Any word or ngram based nearest-neighbors similarity algorithm would have faced the same challenge of unseen actions. On the other hand, DSSM based cosine similarity has been widely used for various semantic similarity based tasks (\cite{ye2016enhancing,palangi2014semantic}), and helps us to map complex queries to their similar queries.  
\end{itemize}

However, finding the similarity of the input query with all the other queries, which are increasing in number, adds latency and is not feasible in real-time systems like personal assistants. Therefore, we use KDTrees-based ANNs of the input query. ANNs yield reasonable results while meeting the real-time latency constraints.

\vspace{0.2cm}
\begin{table}[H]
\caption{Precision vs Recall trade-off for various similarity thresholds in semantic query matching scores.}
\label{tab:threshold} 
\centering
\begin{tabular}{c|c|c|c|c}
\textbf{Similarity} && & &\\
\textbf{Threshold} && \textbf{Precision} & \textbf{Recall} & \textbf{F1 Score} \\
\hline
0.85 && 0.924 & 0.633 & 0.751 \\
0.83 && 0.904 & 0.679 & 0.776 \\
0.82 && 0.896 & 0.699 & 0.785 \\
0.80 && 0.878 & 0.728 & 0.796 \\
0.75 && 0.843 & 0.762 & \textbf{0.801} \\
0.70 && 0.752 & 0.791 & 0.771 \\
0.65 && 0.712 & 0.829 & 0.766 \\
0.60 && 0.667 & 0.850 & 0.745 \\
\end{tabular}
\end{table}

The trade-off between precision and recall for various similarity thresholds on Validation set is shown in Table \ref{tab:threshold}. As it can be seen, reducing the threshold yields higher recall at the cost of lower precision and vice-versa. We chose the similarity threshold to be greater than 0.75, as it gives the best F1-score of 0.801.

Table \ref{tab:comp} compares the performance of both the approaches on Test set for response fetching. DSSM based ANN approach yields higher F1 score.

\vspace{0.2cm}
\begin{table}[H]
\caption{Comparison of POS-based and DSSM based response matching approaches.}
\label{tab:comp} 
\centering
\begin{tabular}{c|c|c|c}
\textbf{Module} & \textbf{Precision} & \textbf{Recall} & \textbf{F1 Score}\\
\hline
POS-based action-skill pairs & 0.89 & 0.61 & 0.72 \\
DSSM-based ANNs & 0.81 & 0.74 & \textbf{0.77} \\ 
\end{tabular}
\end{table}

\section{Conclusions and Future Work} \label{conclusions}
In this paper, we introduced the research problem of detecting and understanding help queries in personal assistants. We proposed an interactive conversational help system which detects help queries and returns appropriate responses. We experimented with traditional as well as popular deep learning architectures to classify help queries and found that C-BiLSTM, a fusion of CNN and BiLSTM, yields the best results. We used DSSM-based semantic similarity to find ANNs of the input query and used them to map the input query to an appropriate predefined response. While the use case is specific, the help model is generic and can be used for queries of other domains.

As part of the future work, we would evaluate other semantic query-query and query-response similarity approaches and compare them against DSSM-based similarity. While the system works for skills with lesser number of training samples, we would like to explore models which yield better results and imporve scalability to new skills.


\end{document}